\documentclass[fleqn,12pt,twoside]{article}
\usepackage{espcrc1}

\usepackage{graphicx}
\usepackage[figuresright]{rotating}

\newcommand{\AmS}{{\protect\the\textfont2
  A\kern-.1667em\lower.5ex\hbox{M}\kern-.125emS}}

\title{Polarized Structure Functions in the Valence Quark and Resonance 
Regions and the GDH Sum.}

\author{Z.-E. Meziani\address[MCSD]{Department of Physics, 
        Temple University, \\ 
        Philadelphia, Pennsylvania, 19122, USA}%
        \thanks{For the Hall A Polarized $^3$He Collaboration}.}
       
\begin{document}

% typeset front matter
\maketitle

\begin{abstract}
I present in this paper the neutron spin physics program in Hall A at 
Jefferson Laboratory using a polarized $^3$He target. The program encompasses 
several completed experiments, in which, valuable spin observables (spin dependent 
structure functions) were measured in order to learn about how the nucleon spin 
 arises from the behavior of the constituents. These experiments also offer 
a ground  for testing our understanding of the strong  regime of quantum 
chromodynamics (QCD) the theory of strong interactions through the  determination 
of moments of these structure functions.
\end{abstract}

\section{INTRODUCTION}

Technical advances for producing polarized beams and polarized targets in
the early 80's triggered a novel experimental effort at SLAC which provided
for the first measurement of the proton spin asymmetry
$A_1^p(x,Q^2$)~\cite{Vhg:83}. The limited $x$ range of this
measurement led to an interpretation of the results consistent with the
predictions of a naive non-relativistic constituent quark model (CQM). However after
EMC at CERN \cite{Ash:88} extended the range of the
SLAC measurement to significantly lower values of $x$, a different and surprising 
picture of the spin structure of the nucleon emerged. By combining 
the measured $\Gamma_1=\int_0^1 g_1(x,Q^2) dx$ with the 
neutron and hyperons $\beta$ decays within the CQM it was found that only a small
fraction of the total nucleon spin was accounted for by quarks. Since then, enormous 
theoretical and experimental progress has been achieved in understanding the nucleon spin
structure\cite{Ehg:99,Fil:01}. The original findings of EMC were confirmed and measurements on the neutron allowed
the test of the Bjorken sum rule~\cite{Bjo:66} to $< 10$\% level. This accurate test was only possible
after the sum rule, originally derived  at $Q^2 \rightarrow \infty $ using current algebra, 
was  re-derived using the technique of operator product expansion (OPE)  within
QCD~\cite{Kod:79,Kod:80,Lar:91,Shu:82}. This step was essential in order to calculate  the corrections 
necessary to evolve this sum rule to $Q^2$  values accessible experimentally. In these
studies most of the measurements on the nucleon spin structure functions  were performed at 
large $Q^2$ above 1 GeV$^2$. Nevertheless to take full advantage of the OPE and test its
applicability in the strong regime of QCD, it was understood that measurements of the
nucleon spin structure below $Q^2$ = 1 GeV$^2$ and in the resonance region would be important for determining the
size of  the higher twists corrections. In this connection it was realized that the Bjorken sum rule was only a
special  case of a more general sum rule known as the "extended" Gerasimov-Drell-Hearn (GDH) sum rule which coincides with the
GDH sum rule at  $Q^2 =0$ and with the Bjorken sum rule at $Q^2\rightarrow \infty$.

Furthermore, higher moments of the spin structure functions are
connected with specific twist matrix elements (observables) which can be
evaluated using Lattice QCD~\cite{Fuk:95,Goc:96}. The advantage  of
measuring higher moments of the spin structure functions is twofold, 1) the 
kinematical region which gives most of the contribution to these moments is 
experimentally accessible 2) calculations of specific matrix elements related 
to these moments through OPE are possible using Lattice QCD. 

I will present here results of two experiments performed at Jefferson Lab in Hall A that
address two areas of the above mentionned concerning the spin structure of the neutron. First we present results of a
precision measurement (E99-117) of the neutron asymmetry $A_1^n$ in the valence region  where the 
existing world data could not confirm the expected behavior from either the CQM or pQCD 
due to a poor statistical precision.  Second we present results of a measurement of both $g_1$ and
$g_2$ (E94-010) through the resonance region at a $Q^2$ ranging from 1 GeV/c to 0.1 GeV/c and 
the neutron $\Gamma_1^n$ and $d_2^n$ are evaluated in the same range of $Q^2$ connecting the
perturbative regime to the strong regime of QCD. 

\section{EXPERIMENTAL METHOD AND SETUP}

The experiments described in this presentation share the same
experimental setup~\cite{Exp:02}. They were carried out at Jefferson Lab in Hall A using a highly
polarized  electron  beam (70-80\%) with an average current up to
15$\mu A$ and a high pressure  polarized (on average between 30\%  and  40\% 
in-beam) $^3$He target with the highest polarized luminosity in the world.  The
target is based on the spin  exchange principle where rubidium atoms are 
polarized by optical pumping and 
their polarization is transferred to the $^3$He atoms via spin exchange
collisions~\cite{Wal:97}.  The target was polarized either parallel or
perpendicular to the direction of the electron beam and its polarization
measured by three independent methods: NMR through the Adiabatic Fast
Passage technique, Electron Paramagnetic Resonance technique (EPR) and
elastic scattering off $^3$He. In each case the results of the three methods agreed
within systematic errors. 

The scattered electrons were detected using two HRS spectrometers
which sat  symmetrically on each side of the electron beam 
line to double the count rate.
The spectrometers were equipped with standard detector packages which consists of 
two vertical drift chambers for momentum and scattering angle determination,
a CO$_2$ gas Cherenkov counter and a double-layered
lead-glass shower counter for particle identification~\cite{And:02}. The 
electron identification efficiency was 99\% and  the $\pi^-$ rejection factor was 
found to be better than 10$^4$ for both spectrometers which was sufficient for all
experiments.  The momentum of each spectrometer was stepped through the relevant excitation
spectrum at each incident energy of a given experiment. In many cases the comparison 
of the data between the two identical  spectrometers at symmetric angles
allowed us to minimize many of the systematic uncertainties.

\section{SPIN AND FLAVOR DECOMPOSITION IN THE VALENCE QUARK REGION}

Among the deep inelastic lepton scattering (DIS) spin observables important for 
understanding the structure of the nucleon  at large $x$, the virtual photon-nucleon
asymmetry $A_1^n$ and spin structure function $g_1^n$ are the most poorly known.   
This shortcoming is due to the small scattering cross sections in the large $x$ region combined with
the lack of high polarized luminosity facilities.  In the past experimental effort has gone into
measuring $A_1^n$ and $g_1^n$ at low $x$, since it provides for most of the strength to $\Gamma_1$,
resulting in  a poor statistical precision in the world data for $x>0.4$. This region, however, is
clean and unambiguous since it is not polluted by sea quarks and gluons and thus offers a unique 
opportunity to test predictions that are usually difficult if not impossible at low $x$. 
A very clean  contribution of the "valence quarks"  can  be expected when $x> 0.5 $.  At 
present at large $x$  the world data are still consistent with the most naive of CQM's which uses a
static SU(6) symmetric  wave function to describe the ground state of the nucleon. However, we know 
from the ratio of neutron to proton unpolarized structure functions measurements 
that this symmetry is broken in nature and thus the prediction is unrealistic.

The set of predictions of $A_1^n$ in the valence quark region falls into two
categories, those of CQM's which break SU(6) symmetry in the ground state 
wave function by hyperfine interaction~\cite{Clo:88,Isg:99} which responsible for the 
N-$\Delta$ mass splitting, and those of perturbative QCD with a hadron helicity conservation (HHC)
constraint~\cite{Jac:75,Bro:95} as $x\rightarrow 1$ which break SU(6) symmetry dynamically. While
both classes of models predict that $A_1^n\rightarrow 1$ when $x\rightarrow 1$ their $x$ dependence
is noticeably different, allowing for a softer variation of $A_1^n$ towards  unity in the case of the CQM's. 
In the case where the HHC constraint  is not imposed in pQCD calculations the results are rather a fit to the world
data~\cite{Lea:02}.  In this case the approach is to perform a pQCD analysis of the proton and neutron world  data
of the polarized structure functions at next-to-leading-order (NLO) and produce a global fit  with a {\it minimal
number of parameters}~. This endeavor has been successful in the calculations by Bourelly, Soffer and Bucella
~\cite{Bou:02} using  a statistical physical picture of the nucleon. 

The difference between the approaches described above is more pronounced when the constituents flavor
decomposition is performed. In the case where we consider a proton, we have 
$\Delta u(x)/u(x)\rightarrow 1$, $\Delta d(x)/d(x)\rightarrow 1$, for the case of pQCD with HHC models,
and $\Delta u/u \rightarrow 1$, $\Delta d/d \rightarrow -2/3$ for the case of CQM's. 
We notice that in the pQCD with HHC models  $\Delta d/d$ changes sign from negative at low $x$ 
to positive at large $x$. This behavior is clearly different in the CQM model based on single 
flavor dominance where $\Delta d/d$ is negative at all $x$ values.

Furthermore, we mention exceptions to the models discussed  above. A bosonized chiral quark model where 
the nucleon appears as a soliton in the meson fields was used to calculate $A_1^n$ in~\cite{Gam:97}.
 In~\cite{Mel:01} local quark-hadron duality is used  to link the structure functions $g_1$ and $F_1$ at large 
$x$  with the $Q^2$ dependence of the nucleon electromagnetic form factors. A prediction was possible however 
we do not know how low in $x$ will this prediction works. Finally structure functions have been calculated in 
the bag model and the effect of the pion cloud on these functions was evaluated in~\cite{Sch:91}.

At Jefferson Lab we are in a unique position to take advantage of the unprecedented polarized luminosity.
The experiment was carried  the highest available incident energy of 5.7 GeV. Both the $A_{\parallel}$ and
$A_{\perp}$ asymmetries for $^3$He were extracted after correcting the raw asymmetries by the beam and
target polarizations and the dilution factor at three kinematics: $x$ = 0.331, 0.474 and 0.61, and $Q^2$ = 2.738,
3.567 and 4.887 GeV$^2$, with  $W^2$ = 6.426, 4.846 and 4.023 GeV$^2$ respectively. False asymmetries were
checked by measuring asymmetries of the polarized beam scattering off an unpolarized $^{12}$C
target and found to be negligible. Radiative corrections were applied to obtain the Born
asymmetries and then, the  asymmetry $A_1$ and the ratio of polarized to unpolarized 
structure functions $g_1/F_1$ of $^3$He were extracted using 

\begin{equation} 
A_1 = \frac{A_{\parallel}}{D(1+\eta \xi)} - \frac{\eta A_{\perp}}{d(1+\eta \xi)}~~{\rm and }~~
\frac{g_1}{F_1}= \frac{1}{D'} \biggl [ A_{\parallel} + \tan{(\theta/2)} A_{\perp} \biggr ],
\end{equation} 
where  $D$, $d$, $\eta$ and $\xi$ and $D'$ are kinematical variables. Finally a
$^3$He model which includes the $S$, $S'$ and $D$ states and pre-existing $\Delta$
components in the $^3$He ground state wave function was used to extract $A_1^n$ and  $g_1/F_1$of the
neutron from that of $^3$He~\cite{Bis:01}.

\begin{figure}[htb]
\begin{center}
\includegraphics[scale=0.95]{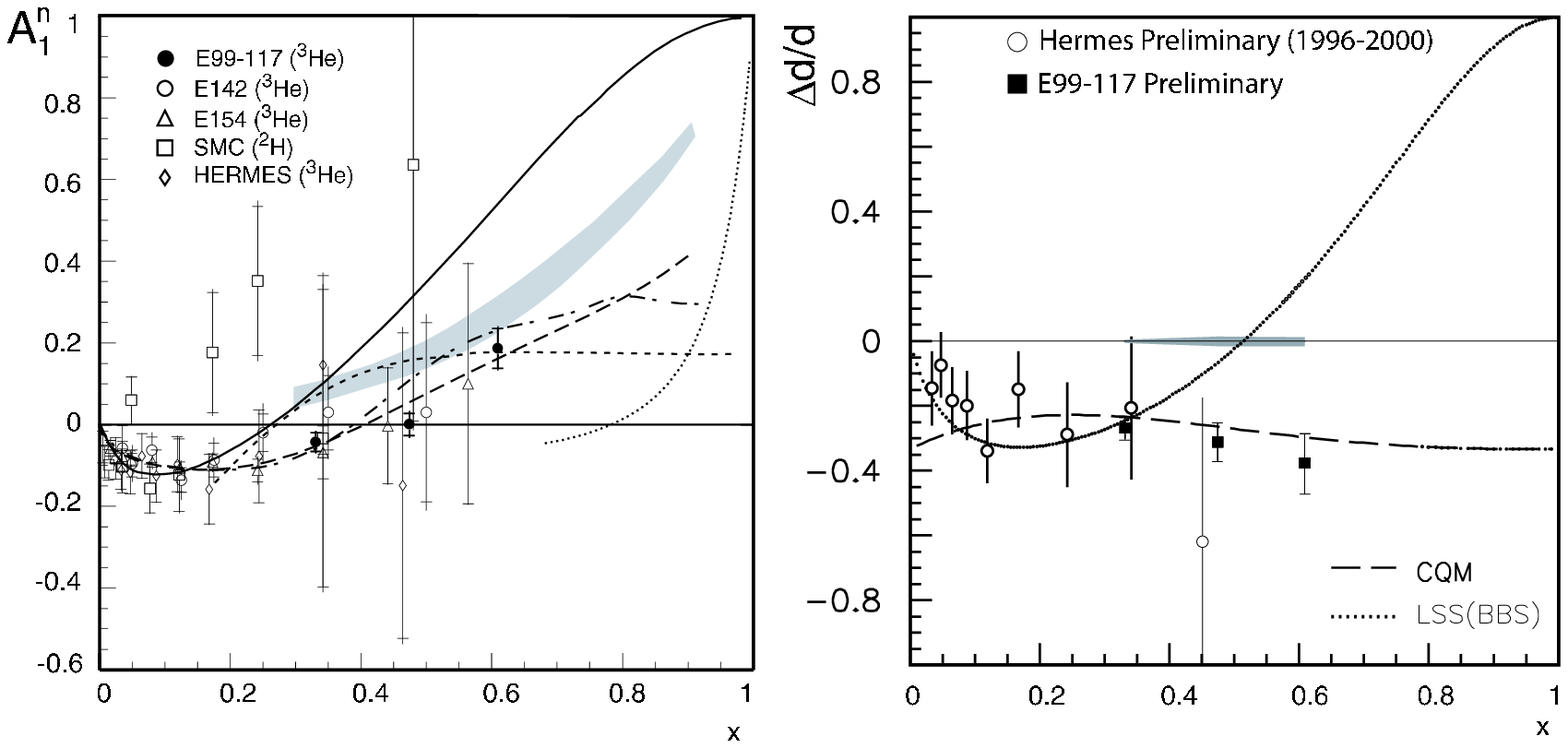}
\caption{Left panel: Preliminary results of Jefferson Lab experiment E99-117 (solid squares)
 along with the world data (open symbols). The curves are predictions described in the
text. Right panel: Spin and flavor dependent quark distributions extracted from the same experiment 
using equation (2). Hermes data error bars are statistical only. Error band shows the uncertainty 
due to neglecting $s$ and $\bar s$.}
\end{center}
\label{fig:a1n}
\end{figure}

In Figure 1 (right panel) we show preliminary results of $A_1^n$. The first data
point at $x=0.33$ is in good agreement with previous measurements. The data points show a clear change of sign 
of $A_1^n$ as $x$ increases and are compared with theoretical predictions. The total error in each
point is dominated by the statistical error. The solid line is a pQCD HHC based on LSS(BBS) parameterization of
${g_1^n}/{F_1^n}$~\cite{Lea:98}, the long-dashed line is a prediction of ${g_1^n}/{F_1^n}$ from LSS 2001
parametrization at Q$^2$ = 5 GeV$^2$ without  HHC constraints~\cite{Lea:02}. The shaded area is a
range of predictions of $A_1^n$ from the constituent quark model~\cite{Isg:99} while the
dot-dashed line  is a calculation of the statistical model at Q$^2$ = 4 GeV$^2$ by
 Bourrely {\it et. al.}~\cite{Bou:02}, and the dotted line is the local duality prediction by 
Melnitchouk~\cite{Mel:01}. Finally,  the short dashed line is the chiral soliton model prediction at Q$^2$ = 3
GeV$^2$ by Weigel, Gamberg and Reinhardt~\cite{Gam:97}. Data from Hermes and SLAC are original values without 
being re-analyzed for the $\Delta$ contribution of the nuclear corrections.

Assuming that the strange quark distributions are negligible in the region $0.3<x<1$ we 
used the quark parton model interpretation of  $g_1$ and  $F_1$ to perform a flavor decomposition of the spin
dependent quark distributions.
 
\begin{equation}
\frac{\Delta u}{u} = \frac {4}{15} \frac{g_1^p}{F_1^p} \biggl (4+
\frac{d}{u} \biggr ) -  \frac {1}{15} \frac{g_1^n}{F_1^n} \biggl (1 + 4\frac{d}{u} \biggr
),~~~~\frac{\Delta d}{d} = \frac {4}{15} \frac{g_1^n}{F_1^n} \biggl (4+
\frac{u}{d} \biggr ) -  \frac {1}{15} \frac{g_1^p}{F_1^p} \biggl (1 + 4\frac{u}{d} \biggr ). 
\end{equation}
Here $u$ and $d$ represent the sum of quark and antiquark distributions.
Using equation (2) and the $d/u$ ratio extracted from the proton and
deuteron structure functions data~\cite{Mel:96} we present in Fig.~1~(left panel) results of 
the down quark distributions obtained in E99-117 (filled squares) along with preliminary results of the HERMES
semi-inclusive measurements (open circles)~\cite{Wen:02}. The solid line is a pQCD fit
to the world data using the HHC constraint as $x\rightarrow 1$. The  dashed line
correspond to a non relativistic constituent quark prediction. It is clear that up to $x = 0.6$ the data favor 
the CQM and is in violation with the HHC pQCD based calculations. We point out that this is consistent
with the interpretation the recent $Q^2$ dependence of the proton electromagnetic form factors
ratio measured at Jefferson Lab~\cite{Kes:02}.

\section{EXTENDED GDH SUM RULE AND TWIST-THREE MATRIX ELEMENT}

In its extended form the GDH sum rule spans all range of momentum transfers and
reads~\cite{Ji:00}:

\begin{equation}
4\int_{Q^2/2M}^\infty {g_1(\nu,Q^2)\over M\nu^2} d\nu  = S_1(Q^2),
\label{eq:gdhsr}
\end{equation}
\noindent where $g_1(\nu,Q^2)$ is the spin structure function of the nucleon, $S_1$ the 
virtual Compton scattering amplitude, $\nu$ the energy transfer and $M$ the nucleon
mass. Starting the integration in equation (3) from the pion production threshold this sum rule can be expanded around $Q^2=0$ 
using for example the heavy baryon chiral perturbation theory (HB$\chi$PT)~\cite{Ji:01}.

\begin{eqnarray}
\Gamma_1^n(Q^2) &=&  -{Q^2\over 8M^2}\kappa_n^2 + 
\frac{5.54}{2M^2}Q^4 +\alpha \biggl ({Q^2\over M^2}\biggl )^3 + ...
\end{eqnarray}
where $\kappa_n$ is the anomalous magnetic moment of the neutron. 
The slope of the first term in the right hand side of equation (4) corresponds to the standard GDH
sum rule~\cite{Ger:65} while the  second term on the right hand side has been evaluated using HB$\chi$PT. Other 
evaluations are available~\cite{Ber:93,Ber:02} and will be compared to the data below.
\noindent In the large $Q^2$ region Eq.~(\ref{eq:gdhsr}) takes another form. If we consider
three quarks flavors and three-loop result for the twist-two component $\Gamma_1^n(Q^2) $is given
by~\cite{Ji:97}:
\begin{eqnarray}
\Gamma_1^n(Q^2) &=& \sum_{i=2,4...}{{\mu_i}(Q^2)\over (Q^2)^{i-2\over 2}}=\mu_2^n (Q^2) + \frac{\mu_4^n (Q^2)}{Q^2}
+ ... \nonumber\\ 
& = &\Big [1-({\alpha_s\over \pi}) - 3.5833({\alpha_s\over \pi})^2 - 20.2153 ({\alpha_s\over \pi})^3
\Big ]  
 \biggl (-{1\over 12} g_A + {1\over 36} a_8 \biggr )    \nonumber \\
& +&  \Big [1 - 0.3333({\alpha_s\over \pi}) - 0.54959({\alpha_s\over \pi})^2 - 4.44725({\alpha_s\over \pi})^3\Big ]~{1\over
9}\Sigma_{\infty} +...
\end{eqnarray}
where $g_a=1.257$ and $a_8=0.579$~\cite{Clo:94} are the triplet and octet axial charge, respectively.
$\Sigma_{\infty}$  is defined as the renormalization group invariant nucleon matrix element of the
singlet  axial current~\cite{Lar:94}. At large enough $Q^2$ the $\mu_4$ contribution becomes
negligible. 

Presently, there is a large set of data on the spin structure function at $Q^2\ge$ 1 GeV$^2$ used to
determine $\Sigma_{\infty}$ and $\alpha_s$. However, below $Q^2 = 1$~GeV$^2$ the experimental situation is less than
ideal, prompting for the investigation of higher twists as $Q^2$ decreases below 1 GeV$^2$.  
This experimental situation has now changed with the completion of experiment E94-010 which covers a Q$^2$ ranging from
0.1~GeV$^2$ to 1~GeV$^2$. In this experiment we have measured the helicity dependent electron scattering cross
sections for excitations energies covering  from the quasielastic region up to the deep inelastic region, thus
including the resonance region.  The longitudinal ($\Delta\sigma_\|$)  and  transverse ($\Delta\sigma_\bot$) cross
section differences are formed from combining data taken with opposite beam helicity,
\begin{equation}
\Delta\sigma_\| = {\sigma^{\downarrow\Uparrow}-\sigma^{\uparrow\Uparrow}},~~~~~~
\Delta\sigma_\bot = {\sigma^{\downarrow\Rightarrow}-\sigma^{\uparrow\Rightarrow}}, 
\label{eq:rawasym}
\end{equation}
where ${\sigma^{\downarrow\Uparrow(\uparrow\Uparrow)}}$ and 
${\sigma^{\downarrow\Rightarrow(\uparrow\Rightarrow)}}$ are spin dependent 
inclusive differential cross sections with the electron spin helicity anti-parallel (parallel) and 
anti-perpendicular (perpendicular) to the target spin direction. The $^3$He spin structure functions $g_1(x,Q^2)$ and 
$g_2(x,Q^2)$ were extracted from the expression:

\begin{eqnarray}
   g_1(x,Q^2) &=& {MQ^2\nu \over 4\alpha_e^2}{ E\over E'}{1\over E+E'} \biggl [\Delta\sigma_\|+
\tan(\theta/2)\Delta\sigma_\bot\biggr ] \\
   g_2(x,Q^2) &=& {MQ^2\nu^2 \over 4\alpha_e^2}{1\over 2E'(E+E')} \biggl [-\Delta\sigma_\| +
\frac{E+E'\cos{\theta}}{E'\sin{\theta}}\Delta\sigma_\bot\biggr ]
\label{eq:g1g2}
\end{eqnarray}

\noindent where $E$ and $E'$ the incident and the scattered electron energy respectively,
$\theta$ is the scattering angle, $M$ is the nucleon mass  and $\alpha_e$ is the
electromagnetic fine coupling constant.

\begin{figure}[h]
\begin{center}
\includegraphics[scale=0.9]{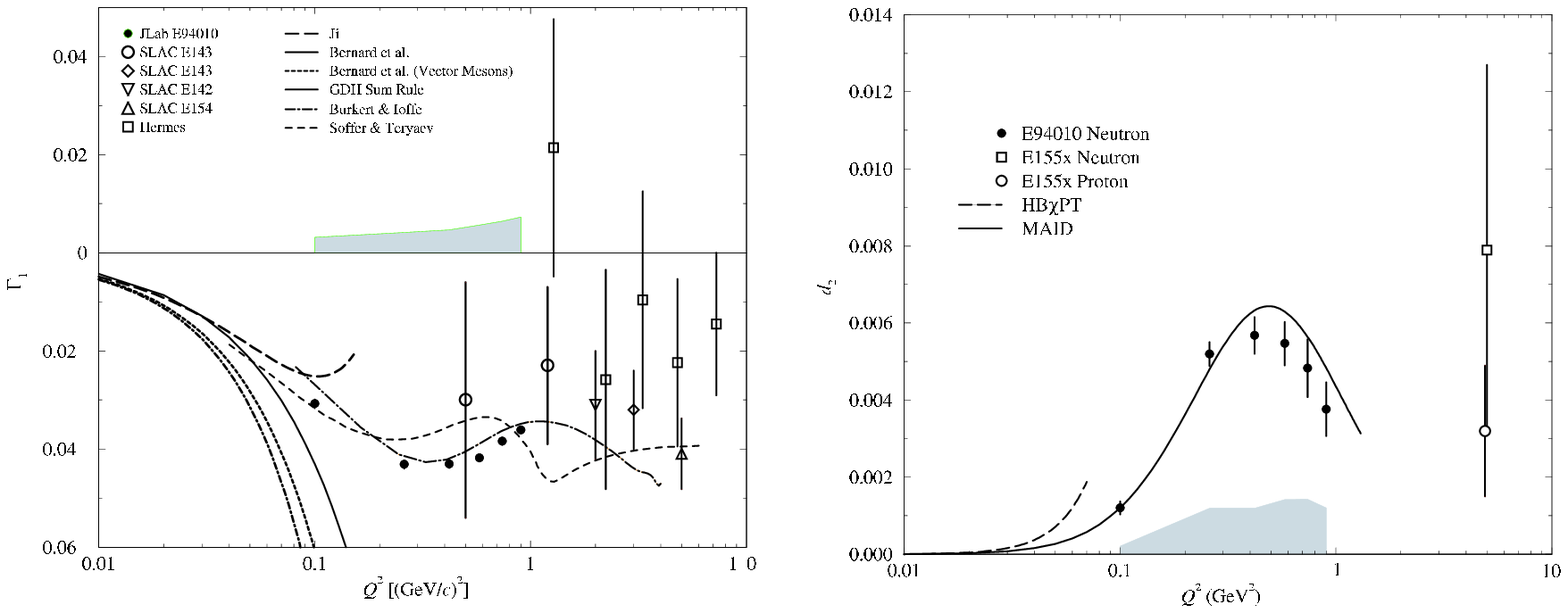}
\caption{Left panel: Preliminary results of the $Q^2$ evolution of $\Gamma_1^n$ from JLab experiment E94-010
along with the world data from DIS and theoretical calculations (see text). Right panel:  Preliminary
results of the twist three matrix element $d_2^n$ compared to SLAC E155X results and theoretical predictions.
(see text). 
\label{fig:gdh_int}}
\end{center}
\end{figure}

The integral $\Gamma_1(Q^2)$ for the neutron is evaluated by first performing the
integration  of the $^3$He response at constant $Q^2$ from the nucleon single pion production
threshold to the lowest $x$ measured in the experiment. We then subtracted a small
contribution due to the quasielastic tail  and applied nuclear corrections to the integral
following the procedure described in ref.~\cite{Cio:97}. We estimated the missing deep
inelastic contribution in the range 2~GeV $<W<$ 30~GeV according to the parametrization of
Thomas and Bianchi  parametrization~\cite{Tho:00}. 

In Fig.~\ref{fig:gdh_int} (left panel), we show $\Gamma_1^n(Q^2)$  at five values of $Q^2$ from 0.1 to about 1 GeV$^2$ 
by step of 0.2 GeV$^2$ (solid circles) compared to the world data (open symbols). The band located  around zero represents the
size of systematic errors. At low $Q^2$ the lines correspond to $\chi$PT calculations by Bernard et. al.~\cite{Ber:93} (short dashed-dot
line) without vector mesons  and  by Bernard et. al.~\cite{Ber:02} (dotted line) with vector
mesons, respectively.  The long dashed line is a calculation by Ji et. al.~\cite{Ji:00,Ji:01} 
 using heavy baryon $\chi$PT.  The solid black line represents the GDH result. The comparison with our data 
shows that beyond $Q^2$=0.1 GeV$^2$  these calculations do not reproduce the trend of the data  because the $\Delta$ degrees 
of freedom  might become important but are not included in both  $\chi$PT calculations.
At moderate and large $Q^2$ two calculations, one by Soffer and Terayev~\cite{Sof:02} (short dashed
curve), the other by Burkert and Ioffe~\cite{Bur:92} (long dash-dot line) are plotted. Soffer and Terayev assume
that the integral over $g_T= g_1 + g_2$ varies smoothly from high $Q^2$ where $g_2\approx 0$ down to $Q^2=0$. 
Using their  prediction for this integral and subtracting the contribution from $g_2$ using the Burkhardt
Cottingham sum rule~\cite{Buc:70}  gives the short dashed curve in Fig.~\ref{fig:gdh_int}, which agree relatively
well with the data.  Burkert and Ioffe consider the contributions from the resonances using the code AO, and the 
nonresonant contributions using a simple higher-twist type form fitted to the  deep-inelastic data. Their model is
constrained to fit both the GDH and the  deep-inelastic limits, and it describes the data quite well.

In the high $Q^2$ regime, using OPE, the $d_2$ matrix element defined as
\begin{equation}
d_2(Q^2) = \int_0^1 x^2[2g_1(x,Q^2)+3g_2(x,Q^2)] dx,
\end{equation}
is shown to be a measure of the electric and magnetic polarizabilities of the color field~\cite{Fil:01}. 
At small $Q^2$, a region covered by our data, its conventional
interpretation in terms of higher twist is not obvious. A recent
publication~\cite{Van:02} gives a new insight in this regard at low $Q^2$ using rather $\chi$PT. Thus, the 
$Q^2$ evolution of $d_2$ is also a quantity that can shed some light on the strong interaction in the nucleon.

In Fig.~\ref{fig:gdh_int} (right panel), the $d_2$ matrix element is shown at several values of $Q^2$ where the integration 
in equation (9) excludes the elastic peak.
The results of this experiment are the solid circles and the grey band represents their corresponding systematic uncertainty.
The SLAC E155~\protect\cite{Ant:02} proton (open circle)  and neutron (open square) results are also shown. The solid line
is the MAID calculation\protect\cite{Dre:01} while the dashed line is a Heavy  Baryon $\chi$PT
calculation\protect\cite{Van:02} valid only a very low $Q^2$. The Lattice prediction~\cite{Goc:96} at $Q^2$ = 5 GeV$^2$
 for the neutron $d_2$ matrix element ( not shown here ) is negative but close to zero. We note that all models predict
$d_2^n$  to be negative or zero at large $Q^2$. At moderate $Q^2$ the data of E94-010 show a positive $d_2^n$  but decreasing
perhaps to zero at high $Q^2$.  The SLAC data also show a positive $d_2^n$ value but with a rather large error bar. More
measurements are needed to  have a complete determination of the transition from low to very high $Q^2$ of this important
matrix element. 

\section{CONCLUSION}
In summary, we presented results of two experiments at Jefferson Lab Hall A which took
advantage of the highly polarized beam and high pressure polarized $^3$He target to investigate the 
internal spin structure of the neutron in the perturbative and the strong regimes of QCD.
In E99-117 we have determined the world most precise spin and flavor dependent distribution 
in the valence region. The results are more in line with the constituent quark model than with the 
HHC constrained pQCD prediction. In experiment E94-010 we have measured both $g_1$ and $g_2$ in the resonance region
allowing  for the first time to study the neutron integrals $\Gamma_1^n$ and the twist three matrix element $d_2^n$ from the pQCD
regime to a regime where chiral perturbation theory might give us more insight into the structure of the nucleon. 
With time we expect Lattice QCD to make predictions in the intermediate region.

\section{ACKNOWLEDGMENTS}
The work presented here was supported in part with funds provided to the Nuclear and Particle
Group at Temple University by the U.S. Department of Energy (DOE) under contract number
DE-FG02-94ER40844. The Southeastern Universities Research Association operates the Thomas Jefferson
National Accelerator  Facility for the DOE under contract DE-AC05-84ER40150

\end{document}